\documentclass[amsmath,amssymb]{revtex4}
\usepackage{graphicx}
\usepackage{bm}

\begin{document}
\title{Scissors mode of trapped dipolar gases}

\author{Misturu Tohyama}
\affiliation{Kyorin University School of Medicine, Mitaka, Tokyo
  181-8611, Japan} 
\begin{abstract}
We study the scissors modes of dipolar boson and fermion gases trapped 
in a spherically symmetric potential. We use
the harmonic oscillator states to solve
the time-dependent Gross-Pitaevskii equation for bosons and the time-dependent Hartree-Fock equation for
fermions. It is pointed out that the scissors modes of bosons and fermions can be of quite different nature.
\end{abstract}
\maketitle
\section{Introduction}
\begin{figure}[h] 
\begin{center} 
\includegraphics[height=5cm]{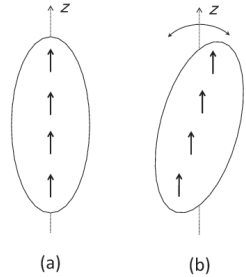}
\end{center}
\caption{Schematic illustration of a new kind of the scissors mode. The arrows pointing the $z$ axis indicate
the polarized dipoles. Since a tilted configuration (b) is energetically 
unfavorable, a dipolar restoring force acts to return the gas to an aligned configuration (a).} 
\label{fig0} 
\end{figure}

An oscillation of a trapped ultracold gas, the so-called scissors mode, has theoretically been
investigated by Gu$\acute{\rm e}$ry-Odelin and Stringari \cite{guery}
in connection to superfluidity associated with Bose-Einstein condensation. 
The mode 
has experimentally been observed in boson gases \cite{marago,cozzini} and also in fermion gases \cite{wright}.
The restoring force of the scissors mode is originated in the deformation of a trapping potential \cite{guery}.
Trapped ultracold dipolar gases, which have recently been attracting experimental and theoretical interests
\cite{Lahaye,Lahaye1,Ni,ospel,Baro}, may provide us with opportunities to study 
a new kind of the scissors mode \cite{bijnen} where the dipolar interaction
plays an important role as a restoring force.
A schematic illustration of such a scissors mode is shown in Fig. \ref{fig0} \cite{bijnen}. 
Since a tilted configuration (b) is energetically 
unfavorable in a completely polarized dipolar gas, a dipolar restoring force acts to bring the gas to an  
aligned configuration (a).
The aim of this paper is to study the properties of such a scissors mode in a dipolar boson gas
and also in a dipolar fermion gas. 
We use the time-dependent mean-field approaches, the time-dependent Gross-Pitaevskii (TDGP) equation for
a boson gas and the time-dependent Hartree-Fock (TDHF) equation for a fermion gas. We
consider gases trapped in a spherically symmetric harmonic potential so that 
pure effects of the dipolar interaction on the scissors mode can be elucidated. To solve these equations, 
we use the harmonic oscillator states with the same frequency as the trapping potential, which facilitates the 
calculation of the matrix elements of the dipolar interaction
and allows us to treat boson and fermion gases in a similar framework. 
A disadvantage of our approach is that it is restricted to weak interaction regime and to rather small
fermion systems.
The paper is organized as follows. The formulation is given in Sect. II, the results are presented 
in Sect. III and Sect. IV is devoted to summary. 

\section{Formulation}
\subsection{Hamiltonian}
We consider trapped dipolar gases of bosons and single-component fermions at zero temperature. 
We assume that the gases are trapped in a spherically symmetric harmonic 
potential with
frequency $\omega$ and that the dipoles are completely polarized
along the $z$ axis. The systems can be described by the hamiltonian 
\begin{eqnarray}
H=\sum_\alpha\epsilon_\alpha a^\dag_\alpha a_\alpha
+\frac{1}{2}\sum_{\alpha\beta\alpha'\beta'}\langle\alpha\beta|v|\alpha'\beta'\rangle
a^\dag_{\alpha}a^\dag_\beta a_{\beta'}a_{\alpha'},
\label{totalH}
\end{eqnarray}
where $a^\dag_\alpha$ and $a_\alpha$ are the creation and annihilation operators of 
a harmonic oscillator state $\alpha$ corresponding to the trapping potential,
$\epsilon_\alpha=\omega(n+3/2)$ with $n=0,~1,~2,....$,
and $\langle\alpha\beta|v|\alpha'\beta'\rangle$ is the matrix element of 
the dipolar interaction $v({\bm r})=(d^2/r^3)(1-3z^2/r^2)$. Here $d$ is the magnetic dipole moment for a magnetic dipole.
For an electric dipole, $d$ is given by $\alpha(0){\cal E}$ where $\alpha(0)$ is the atomic polarizability and ${\cal E}$
the electric-field strength \cite{Yi}. 
We use the units such that $\hbar=1$.
Assuming that it would be possible to achieve zero scattering length through a Feshbach resonance
\cite{werner,pavlo},  we neglect
the contact interaction described by the $s$-wave scattering length, as has often been done 
in theoretical investigations of dipolar boson gases \cite{santos,ronen1}.
In the case of a single-component
fermion gas
the contact term does not contribute because of Fermi statistics \cite{Baro}.
The harmonic oscillator states with a definite $n$ have degeneracy of $(n+1)(n+2)/2$ - fold and are called
an oscillator shell \cite{Fetter}. These degenerate levels contain states with different orbital angular
momentum $l$ and its $z$ component $m_z$.

\subsection{Time-dependent Gross-Pitaevskii equation}
The TDGP equation
for the condensate wavefunction $\Phi_0({\bm r},t)$ is given by \cite{Baro}
\begin{eqnarray}
i\frac{\partial}{\partial t}\Phi_0({\bm r},t)=
 \left[-\frac{\nabla^2}{2m}+V_{\rm ext}
({\bm r})
+\int v({\bm r}-{\bm r'})n({\bm r'},t)d{\bm r'}\right]\Phi_0({\bm r},t),
\label{GP}
\end{eqnarray}
where $m$ is the mass of a boson, $V_{\rm ext}({\bm r})=m\omega^2(x^2+y^2+z^2)/2$ and $n({\bm r},t)=|\Phi_0({\bm r},t)|^2$.
The wavefunction $\Phi_0({\bm r},t)$ is normalized to 
$N$, the number of bosons. We expand $\Phi_0({\bm r},t)$
using the time-independent harmonic oscillator states as
\begin{eqnarray}
\Phi_0({\bm r},t)=\sum_\alpha U_\alpha(t)\phi_\alpha({\bm r}),
\end{eqnarray}
where $\phi_\alpha({\bm r})$ is the harmonic oscillator wavefunction.
In order to treat boson and fermion gases in a similar framework, we derive 
an equation of motion for 
a matrix $N_{\alpha\alpha'}$ defined by
\begin{eqnarray}
N_{\alpha\alpha'}(t)&=&U_\alpha(t)U_{\alpha'}(t)^*.
\end{eqnarray}
The matrix  $N_{\alpha\alpha'}$ satisfies $\sum_\alpha N_{\alpha\alpha}=N$.
Using Eq. (\ref{GP}) and its complex conjugate, we obtain 
the equation of motion for $N_{\alpha\alpha'}$ as \cite{adiabatic1}
\begin{eqnarray}
i\dot{N}_{\alpha\alpha'}&=&
\sum_{\lambda}(\epsilon_{\alpha\lambda}{N}_{\lambda\alpha'}-{N}_{\alpha\lambda}\epsilon_{\lambda\alpha'}),
\label{bose}
\end{eqnarray}
where $\epsilon_{\alpha\alpha'}$ is given by
\begin{eqnarray}
\epsilon_{\alpha\alpha'}=\epsilon_\alpha\delta_{\alpha\alpha'}
+\sum_{\lambda_1\lambda_2}
\langle\alpha\lambda_1|v|\alpha'\lambda_2\rangle
N_{\lambda_2\lambda_1}.
\end{eqnarray}
Equation (\ref{bose}) can also be derived from the hamiltonian Eq. (\ref{totalH}) using the time-dependent
total wavefunction $|\Phi(t)\rangle=\exp [-iHt]|\Phi(t=0) \rangle$ and defining $N_{\alpha\alpha'}$ by
\begin{eqnarray}
N_{\alpha\alpha'}(t)=\langle\Phi(t)|a^\dag_{\alpha'}a_\alpha|\Phi(t)\rangle.
\end{eqnarray}
The time derivative of $N_{\alpha\alpha'}$ is given by
\begin{eqnarray}
i\dot{N}_{\alpha\alpha'}=\langle\Phi(t)|[a^\dag_{\alpha'}a_\alpha,H]|\Phi(t)\rangle.
\end{eqnarray}
Using the Hartree approximation for the expectation values of two-body
operators on the right-hand side of the above equation, 
\begin{eqnarray}
\langle\Phi(t)|a^\dag_{\alpha'}a^\dag_{\beta'}a_\beta a_\alpha|\Phi(t)\rangle\approx N_{\alpha\alpha'}N_{\beta\beta'},
\end{eqnarray}
we obtain Eq. (\ref{bose}).
Equation (\ref{bose}) conserves 
the total energy $E_{0}$ of the condensate given by
\begin{eqnarray}
E_{0}&=&\sum_{\alpha}\epsilon_\alpha N_{\alpha\alpha}
\nonumber \\
&+&\frac{1}{2}\sum_{\alpha\beta\alpha'\beta'}
\langle\alpha\beta|v|\alpha'\beta'\rangle N_{\alpha'\alpha}N_{\beta'\beta}.
\end{eqnarray}
Equation (\ref{bose}) also conserves the total number of bosons. The condensate is given as a stationary solution
of Eq. (\ref{bose}), which satisfies
\begin{eqnarray}
\sum_{\lambda}(\epsilon_{\alpha\lambda}{N}_{\lambda\alpha'}-{N}_{\alpha\lambda}\epsilon_{\lambda\alpha'})=0.
\label{bose1}
\end{eqnarray}
Equation (\ref{bose1}) implies that both $\epsilon_{\alpha\alpha'}$ and $N_{\alpha\alpha'}$ 
can simultaneously be diagonalized. An eigenstate of $N_{\alpha\alpha'}$ which has eigenvalue $N$ is the condensate.

\subsection{Time-dependent Hartree-Fock equation}
The TDHF equation for a fermion gas can be written in a form similar to Eq. (\ref{bose}) 
using the occupation matrix $n_{\alpha\alpha'}$ defined by
\begin{eqnarray}
n_{\alpha\alpha'}(t)&=&\langle\Psi(t)|a^\dag_{\alpha'} a_\alpha|\Psi(t)\rangle,
\end{eqnarray}
where $|\Psi(t)\rangle$ is the time-dependent total wavefunction.
The equation for $n_{\alpha\alpha'}$ is given as 
\begin{eqnarray}
i \dot{n}_{\alpha\alpha'}&=&\langle\Psi(t)|[a^\dag_{\alpha'}a_\alpha,H]|\Psi(t)\rangle.
\end{eqnarray}
Using the Hartree-Fock (HF) approximation for the expectation values of two-body
operators on the right-hand side of the above equation, 
\begin{eqnarray}
\langle\Psi(t)|a^\dag_{\alpha'}a^\dag_{\beta'}a_\beta a_\alpha|\Psi(t)\rangle\approx n_{\alpha\alpha'}n_{\beta\beta'}-
n_{\alpha\beta'}n_{\beta\alpha'},
\end{eqnarray}
we obtain the TDHF equation \cite{RS}
\begin{eqnarray}
i \dot{n}_{\alpha\alpha'}=
\sum_{\lambda}(\epsilon_{\alpha\lambda}{n}_{\lambda\alpha'}-{n}_{\alpha\lambda}\epsilon_{\lambda\alpha'}).
\label{fermi}
\end{eqnarray}
The energy matrix $\epsilon_{\alpha\alpha'}$ is given by
\begin{eqnarray}
\epsilon_{\alpha\alpha'}=\epsilon_\alpha\delta_{\alpha\alpha'}
+\sum_{\lambda_1\lambda_2}
\langle\alpha\lambda_1|v|\alpha'\lambda_2\rangle_A 
n_{\lambda_2\lambda_1},
\end{eqnarray}
where the subscript $A$ means that the corresponding matrix is antisymmetrized.
The TDHF equation (\ref{fermi}) satisfies the conservation laws of the total energy and the
total number of particles \cite{RS}.
The total energy $E_{\rm HF}$ in the HF approximation is given by
\begin{eqnarray}
E_{\rm HF}&=&\sum_{\alpha}\epsilon_\alpha n_{\alpha\alpha}
\nonumber \\
&+&\frac{1}{2}\sum_{\alpha\beta\alpha'\beta'}
\langle\alpha\beta|v|\alpha'\beta'\rangle_A n_{\alpha'\alpha}n_{\beta'\beta}.
\end{eqnarray}
The HF ground state is given as a stationary solution
of Eq. (\ref{fermi}), which satisfies
\begin{eqnarray}
\sum_{\lambda}(\epsilon_{\alpha\lambda}{n}_{\lambda\alpha'}-{n}_{\alpha\lambda}\epsilon_{\lambda\alpha'})=0.
\label{fermi1}
\end{eqnarray}
Equation (\ref{fermi1}) implies that both $\epsilon_{\alpha\alpha'}$ and $n_{\alpha\alpha'}$ 
can be diagonalized simultaneously. The eigenvalues of $n_{\alpha\alpha'}$ are either 1 or 0.

\subsection{Scissors mode}
As mentioned above, 
the condensate of a boson gas and the HF ground state of a fermion gas
are given as stationary solutions of Eq. (\ref{bose}) and Eq. (\ref{fermi}), respectively.
We use the following adiabatic method \cite{adiabatic1,Toh09,pfitz} to obtain nearly stationary
solutions of Eqs. (\ref{bose}) and (\ref{fermi}): Starting from a noninteracting configuration,
we solve  Eqs. (\ref{bose}) and (\ref{fermi}) using a temporally increasing interaction 
$v({\bm r})\times t/T$. The noninteracting configuration for a boson gas is the lowest-energy harmonic oscillator
state and that for a fermion gas consists of the harmonic oscillators states filled up to a certain shell determined by
the number of fermions. We must use large $T$ so that
spurious oscillations coming from mixing of excited states are well reduced. 
We found that the value $T=15\pi/\omega$ is sufficiently large to suppress such unphysical contributions.

In experiments \cite{marago,wright} the scissors mode is excited through a sudden rotation of the trapping
potential by a small angle. In our approach this is achieved by using the time-dependent one-body operator
$k\sum_{\alpha\alpha'}\langle\alpha|Q|\alpha'\rangle
a^\dag_{\alpha}a_{\alpha'}\delta(t-T)$ \cite{Toh09}, 
where $Q$ is the angular momentum operator, $Q({\bm r})=L_x=-i(y\partial/\partial z-z\partial/\partial y)$,   
and $k$ is a parameter determining the tilted angle and, therefore, the amplitude of the oscillation.
The matrix elements
of $L_x$ are nonvanishing only between the harmonic oscillator states with the same
$n$ and $l$ and with the difference in $m_z$ by $\pm 1$. The initial condition for 
$N_{\alpha\alpha'}$ at $t=T$ becomes 
\begin{eqnarray}
N_{\alpha\alpha'}(T_+)=
\sum_{\lambda\lambda'}\langle\alpha|e^{-ikQ}|\lambda\rangle
N_{\lambda\lambda'}(T_-)\langle\lambda'|e^{ikQ}|\alpha'\rangle,
\label{N-init}
\end{eqnarray}
where $T_-$ and $T_+$ indicate the times infinitesimally before and after $T$, respectively.
Equation (\ref{N-init}) gives an initial configuration of a gas rotated along the $x$ axis.
As a consequence, the expectation value of $yz$ is nonvanishing at $t=T$. One may think  
that the operator $yz$ can also be used as the excitation operator of the scissors mode. However,
$yz$ is essentially the quadrupole operator $r^2Y_{2\pm 1}(\theta,\phi)$ and
excites a deformation mode different from the scissors mode.
In the case of a fermion gas, $N_{\lambda\lambda'}$ in Eq. (\ref{N-init}) is replaced by $n_{\lambda\lambda'}$.

We study the scissors mode in the small amplitude regime and, therefore,
expand Eq. (\ref{N-init}) in powers of $k$: We take terms up to the fourth order of $k$.
The linearized form of the TDGP equation may be referred to as the Bogoliubov equations \cite{pethick} and 
that of the TDHF theory is the random-phase approximation (RPA) \cite{RS}. In the small amplitude regime the physical
behavior of collective oscillations are usually investigated using 
the strength function $S(E)$ \cite{RS,dalfovo} (the dynamic structure factor \cite{ewsr}) 
for the excitation operator $Q$, which describes 
the transition strength at excitation energy $E$. In our approach, $S(E)$ 
is calculated as \cite{Toh09}
\begin{eqnarray}
S(E)=\frac{1}{k\pi}\int_0^\infty (q(t)-q(T))\sin Et' dt',
\label{strength}
\end{eqnarray}
where  $q(t)=\langle L_x\rangle$ and $t'=t-T$. 
Since the integration in Eq. (\ref{strength}) is performed for a finite interval
in our numerical calculations,
we multiply $q(t)-q(T)$ by a damping factor $\exp(-\Gamma t'/2)$ to suppress spurious oscillations
in $S(E)$.
The strength function satisfies the
following energy-weighted sum rule expressed with the double commutator of 
$H$ and $Q$ \cite{RS,dalfovo,ewsr},
\begin{eqnarray}
 \int_0^\infty E S(E)dE&=&\frac{1}{2}\langle\Phi_0|[Q,[H,Q]]|\Phi_0\rangle,
 \label{ewsr}
\end{eqnarray}
where $|\Phi_0\rangle$ is the ground-state wavefunction.
The right-hand side of the above equation for $H$ in Eq. (\ref{totalH}) and $Q=L_x$
is written as $d^2\langle\Phi_0|W|\Phi_0\rangle/2$, where $W$ is a two-body operator $W=3((z-z')^2-(y-y')^2)/|{\bm r}-{\bm r}'|^5$. 
Thus the value of the energy-weighted sum for the scissors mode 
directly depends on the interaction strength in contrast to other
collective modes such as the breathing mode and the quadrupole mode \cite{ewsr,RS}: For example
the right-hand side of Eq. (\ref{ewsr}) for the $yz$ quadrupole mode excited by $Q=yz$ is given by $\langle\Phi_0|y^2+z^2|\Phi_0\rangle/2m$.

\section{Results}
\subsection{Dipolar boson gas}
\begin{figure} 
\begin{center} 
\includegraphics[height=5cm]{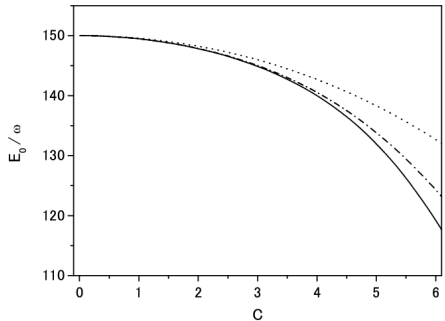}
\end{center}
\caption{Condensate energy for $N=100$ as a function of the dipolar parameter $C=Nd^2/\xi^3\omega$. 
The dotted, dot-dashed and solid line
depict the results obtained using the harmonic oscillator states up to the $n=4$, 6 and 8 shells, respectively.} 
\label{boseeng} 
\end{figure}

\begin{figure} 
\begin{center} 
\includegraphics[height=5cm]{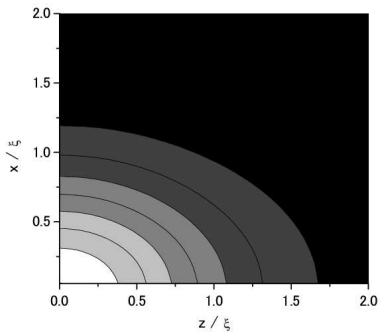}
\end{center}
\caption{Contour plot of the density $n({\bm r})$ in the $x-z$ plane  
for $N=100$ and $C=5$. The white area is the most dense. 
The density has axial symmetry along the $z$ axis and reflection symmetry with respect
to the $x-y$ plane.} 
\label{bosedns} 
\end{figure}

We perform our calculations for a gas of 100 bosons. According to Eq. (\ref{GP}), the obtained results
appropriately normalized by $N$ and $\omega$ depend only on the dipolar parameter $C=Nd^2/\xi^3\omega$,
where $\xi$ is the oscillator length $\xi=\sqrt{1/m\omega}$.
The condensate energy $E_0$ calculated using the oscillator states up to the $n=8$ shell (solid line) 
is shown in Fig. \ref{boseeng} as a function of $C$.
The results obtained using the oscillator states up to the $n=4$ (dotted line) and 6 (dot-dashed line) are also shown 
for comparison. 
More than 93\% 
of the interaction energy $(E_0(C=0)-E_0(C))$ is estimated to be achieved at $C=5$ using the oscillator states
up to the $n=8$ shell. In the following we present the results for $C<5$ obtained using the oscillator states up to the $n=8$ shell.
Our calculations suggest that the dipolar gas is stable in the 
range of the dipolar parameter up until $C\approx5$, which agrees with other
theoretical investigations of
the ground states and excited states of dipolar boson gases 
\cite{ronen1,goral}. 
The energy gain shown in Fig. \ref{boseeng} is achieved by deformation caused by mixing of higher oscillator states.
The condensate has a prolate shape and its deformation increases with increasing interaction strength.
The density profile  at $C=5$ is shown in Fig. \ref{bosedns}.

\begin{figure} 
\begin{center} 
\includegraphics[height=5cm]{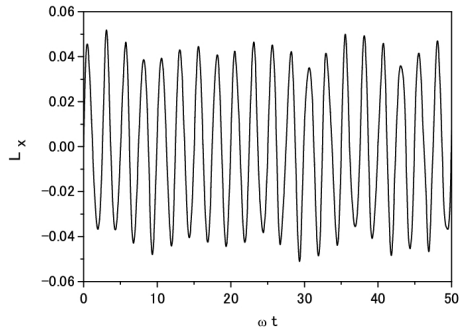}
\end{center}
\caption{Time evolution of $\langle L_x\rangle$ calculated using the TDGP equation for $C=5$ and $k=0.001$.} 
\label{bosefig1} 
\end{figure}

\begin{figure} 
\begin{center} 
\includegraphics[height=5cm]{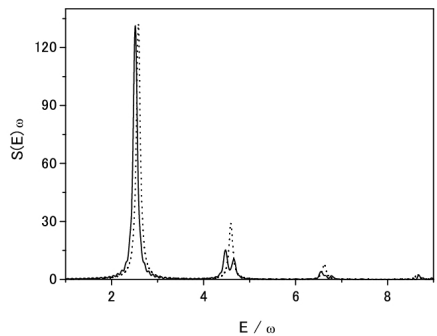}
\end{center}
\caption{Strength function calculated at $C=5$. The dotted line indicates the unperturbed distribution.} 
\label{bosefig2} 
\end{figure}

\begin{figure} 
\begin{center} 
\includegraphics[height=5cm]{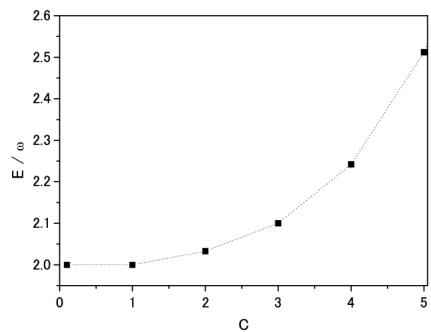}
\end{center}
\caption{Excitation energy of the lowest-energy scissors mode as a function of $C$.} 
\label{bosefig3} 
\end{figure}

The expectation value of $L_x$ is shown in Fig. \ref{bosefig1} as a function of time for $C=5$.
The time evolution slightly differs from a harmonic oscillation, indicating a mixture of different frequencies.
The strength function obtained from Eq. (\ref{strength}) (solid line) is shown in Fig. \ref{bosefig2}.
The artificial width used is
$\Gamma/\omega=0.1$.
The dotted line depicts the unperturbed strength distribution obtained from 
the eigenstates of $N_{\alpha\alpha'}$ (and also $\epsilon_{\alpha\alpha'}$) at $t=T_-$.
The small difference between the two calculations indicates that the condensate-particle correlations such as those  
included in the Bogoliubov equations are small.
The small peaks located at higher energies are originated in the mixing of higher oscillators states into the condensate.
The excitation energy of the lowest-energy peak of the scissors mode
is shown in Fig. \ref{bosefig3} as a function of $C$. 
The excitation energy is $2\omega$ at $C=0$ and increases almost quadratically with increasing $C$.
The fact that the excitation energy is close to $2\omega$ in the small $C$ region means
that the trapping potential plays an essential role as a restoring force in this region.
We found that
the excitation energy of the scissors mode coincides with that of the quadrupole mode excited by the operator $Q=yz$ (or $xz$).
This may be understood by the fact that
major components of both modes consist of the transitions from the condensate to 
the lowest-energy positive-parity states with $m_z=\pm 1$. 
The difference 
is that the scissors mode consists of the transitions with $\Delta l =0$, whereas the $yz$ quadrupole mode can have
the transitions both with $\Delta l =0$ and 2.
The energies of these states with $m_z=\pm 1$ are $2\omega$ higher than
the energy of the condensate at $C=0$ and weakly depend on $C$, while the condensate energy is 
a rapidly decreasing function of $C$ as shown in Fig. \ref{boseeng}. This explains the behavior of the excitation energies of the 
scissors mode (and the quadrupole mode ) shown in Fig. \ref{bosefig3}.

\begin{figure} 
\begin{center} 
\includegraphics[height=5cm]{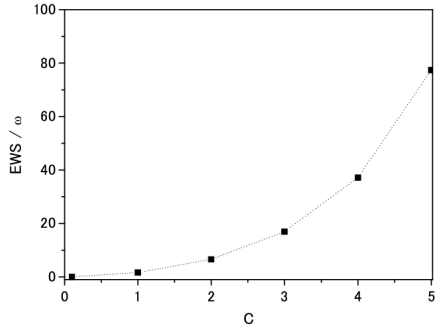}
\end{center}
\caption{Energy-weighted sum of the transition strength as a function of $C$.} 
\label{bosefig4} 
\end{figure}

In the following we discuss the energy-weighted sum of the transition strength.  
The transition strength and its energy-weighted sum give useful information to understand 
the dynamical properties of the condensate \cite{ewsr}, though they may not have experimental significance 
because the scissors mode is excited not by an external field but by a sudden rotation of the trapping potential.
The value of the energy-weighted sum is shown as a function of $C$ in Fig. \ref{bosefig4}. 
As mentioned above, the
sum depends on the interaction strength: It is zero at $C=0$
and increases with increasing $C$. The behavior of the transition strength  
of the scissors mode at small $C$ may be explained by the small mixing of the oscillator state with
$n=2$, $l=2$ and $m_z=0$ into the condensate because the scissors mode is  mainly excited
through the transitions from the component of the oscillator state with
$n=2$, $l=2$ and $m_z=0$ in the condensate to the oscillator states with
$n=2$, $l=2$ and $m_z=\pm 1$. The small mixing of the oscillator state with
$n=2$, $l=2$ and $m_z=0$ in the condensate gives the vanishing transition strength at small $C$.
On the other hand the $yz$ quadrupole mode at small $C$ is mainly excited
by the transitions between the harmonic oscillator state with
$n=0$, $l=0$ and $m_z=0$ in the condensate and the oscillator states with
$n=2$, $l=2$ and $m_z=\pm 1$. Therefore, its
energy-weighted sum, which is given by $\langle\Phi_0|y^2+z^2|\Phi_0\rangle/2m$,
is almost independent of the interaction strength: 
The change in the energy-weighted sum for the $yz$ quadrupole mode is in fact only 10\%
when $C$ is increased to 5. The small transition strength of the scissors mode
implies that the scissors mode may be easily masked by the $yz$ quadrupole mode 
if there is a tiny change in the shape of the trapping potential during
the excitation process of the scissors mode.

\subsection{Dipolar fermion gas}
\begin{figure} 
\begin{center} 
\includegraphics[height=5cm]{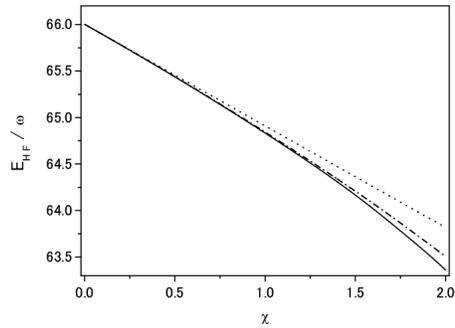}
\end{center}
\caption{Hartree-Fock energy as a function of $\chi=d^2/\xi^3\omega$ for $N=18$. The dotted, dot-dashed and solid line
depict the results obtained using the harmonic oscillator states up to the $n=3$, 5 and 7 shells, respectively.} 
\label{fermieng} 
\end{figure}

\begin{figure} 
\begin{center} 
\includegraphics[height=5cm]{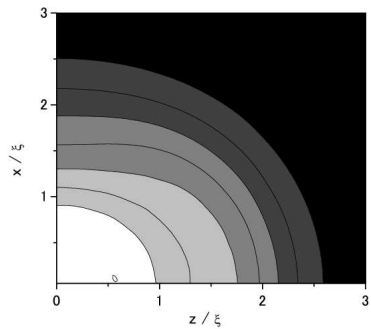}
\end{center}
\caption{Contour plot of the density $n({\bm r})$ in the $x-z$ plane obtained in HF 
for $N=18$ and $\chi=1$. The white area is the most dense. 
The density has axial symmetry along the $z$ axis and reflection symmetry with respect
to the $x-y$ plane.} 
\label{fig1} 
\end{figure}

Since the number of the harmonic oscillator states needed in our approach rapidly increases with increasing
number of fermions, we can only manage rather small systems. 
We mainly consider gases with $N\approx 20$.
Such small $N$ systems have often been
used for theoretical investigations of dipolar fermion gases \cite{Oster}
and may be realized in the array of microtraps or optical lattices as discussed
in Refs. \cite{Oster,Barberan,Popp}.
It is pointed out in Ref. \cite{Toh09} that in a spherically symmetric trapping potential,
dipolar fermion gases with 
the closed-shell configurations where the oscillator states up to a certain oscillator shell are fully occupied   
have small interaction energy and negligible deformation because of the cancelation between the attractive and repulsive parts
of the dipolar interaction. The numbers of fermions corresponding to the closed-shell configurations are
$4$, 10, 20, 35, 56, and so on. Since such closed-shell gases are rather exceptional, 
we first consider a deformed gas 
consisting of 18 fermions, which has 
a simple configuration with two fermions removed from the $N=20$ closed-shell configuration.
Since the dipolar interaction favors a density distribution elongated along the polarized axis,
a configuration where the oscillators states with the highest $|m_z|$ are empty
has the lowest energy among the configurations with $N=18$ as long as the interaction is not strong: 
These vacated states are the oscillator states with $n=3$, $l=3$ and $m_z=\pm 3$.
When the dipole interaction is absent, there is no restoring force for the scissors mode of the gas
because all harmonic oscillator states with the same $n$ and $l$ and with different $m_z$ are degenerate. Therefore, 
pure effects of the dipole interaction can be seen in the scissors mode
of this $N=18$ configuration. 
The HF energy calculated for such a configuration is shown in Fig. \ref{fermieng} as a function of $\chi=d^2/\omega\xi^3$. 
The dotted, dot-dashed and solid line
depict the results obtained using the harmonic oscillator states up to the $n=3$, 5 and 7 shells, respectively.
Since more than 90\% 
of the interaction energy $(E_{HF}(\chi=0)-E_{HF}(\chi=2))$ 
is estimated to be achieved,
we present the results obtained using the oscillator states up to the $n=5$ shell. 
Since the noninteracting state is already deformed, mixing of the oscillator states is small 
in contrast to the boson gas considered above.
The change in the deformation is also small: The quadrupole moment decreases by 5\%
when $\chi$ increases to 2.
As a consequence, the HF energy almost linearly decreases with increasing $\chi$. 
The density profile calculated for $N=18$ and $\chi=1$ is shown in Fig. \ref{fig1}.
 
\begin{figure} 
\begin{center} 
\includegraphics[height=5cm]{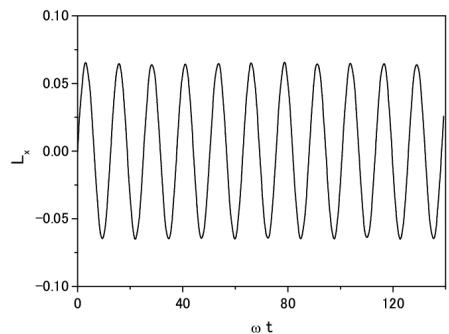}
\end{center}
\caption{Time evolution of $\langle L_x \rangle$ calculated in TDHF for $N=18$, $\chi=1$
and $k=0.01$.} 
\label{fig2} 
\end{figure} 

\begin{figure} 
\begin{center} 
\includegraphics[height=5cm]{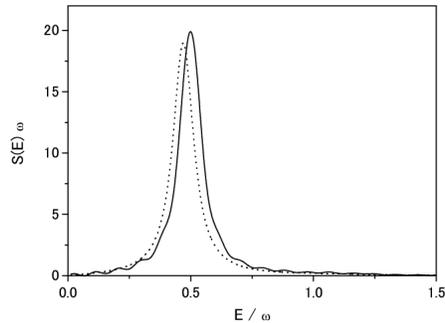}
\end{center}
\caption{Strength function calculated in TDHF for $N=18$ and $\chi=1$. The dotted line depicts the unperturbed distribution.} 
\label{fig3} 
\end{figure}

The time evolution of $\langle L_x \rangle$ at $\chi=1$ is shown in Fig. \ref{fig2}. It shows a sinusoidal
oscillation pattern. The strength distribution obtained from Eq. (\ref{strength}) is 
shown in Fig. \ref{fig3}.   
The unperturbed distribution obtained from the eigenstates of $\epsilon_{\alpha\alpha'}$ at $t=T_-$
is also shown with the dotted line. The difference between the solid line and dotted line is 
attributed to particle - hole correlations. Figure \ref{fig3} indicates that the effects of
particle - hole correlations are quite small. As mentioned above,
the transition induced by the operator $L_x$ is allowed only between the oscillator states 
which have the same $n$ and 
$l$ and differ in $m_z$ by $\pm 1$. In the case of $N=18$ the single-particle transitions from the occupied negative-parity 
states with $m_z=\pm 2$
to the unoccupied negative-parity states with $m_z=\pm 3$ are dominant and the major components of these single-particle states
are the harmonic oscillator states with $l=3$.

\begin{figure} 
\begin{center} 
\includegraphics[height=5cm]{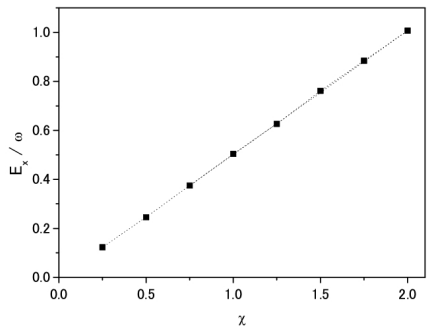}
\end{center}
\caption{Excitation energy as a function of $\chi$ for $N=18$.} 
\label{fig5} 
\end{figure} 

\begin{figure} 
\begin{center} 
\includegraphics[height=5cm]{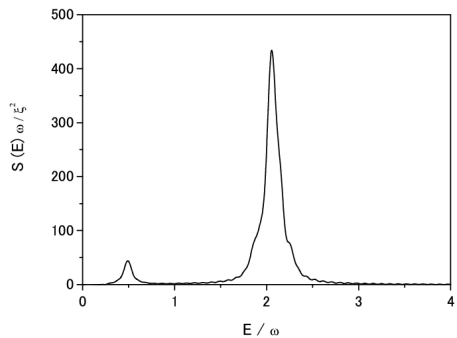}
\end{center}
\caption{Strength function for the $yz$ quadrupole mode calculated in TDHF for $N=18$ and $\chi=1$.} 
\label{yz} 
\end{figure} 

\begin{figure} 
\begin{center} 
\includegraphics[height=5cm]{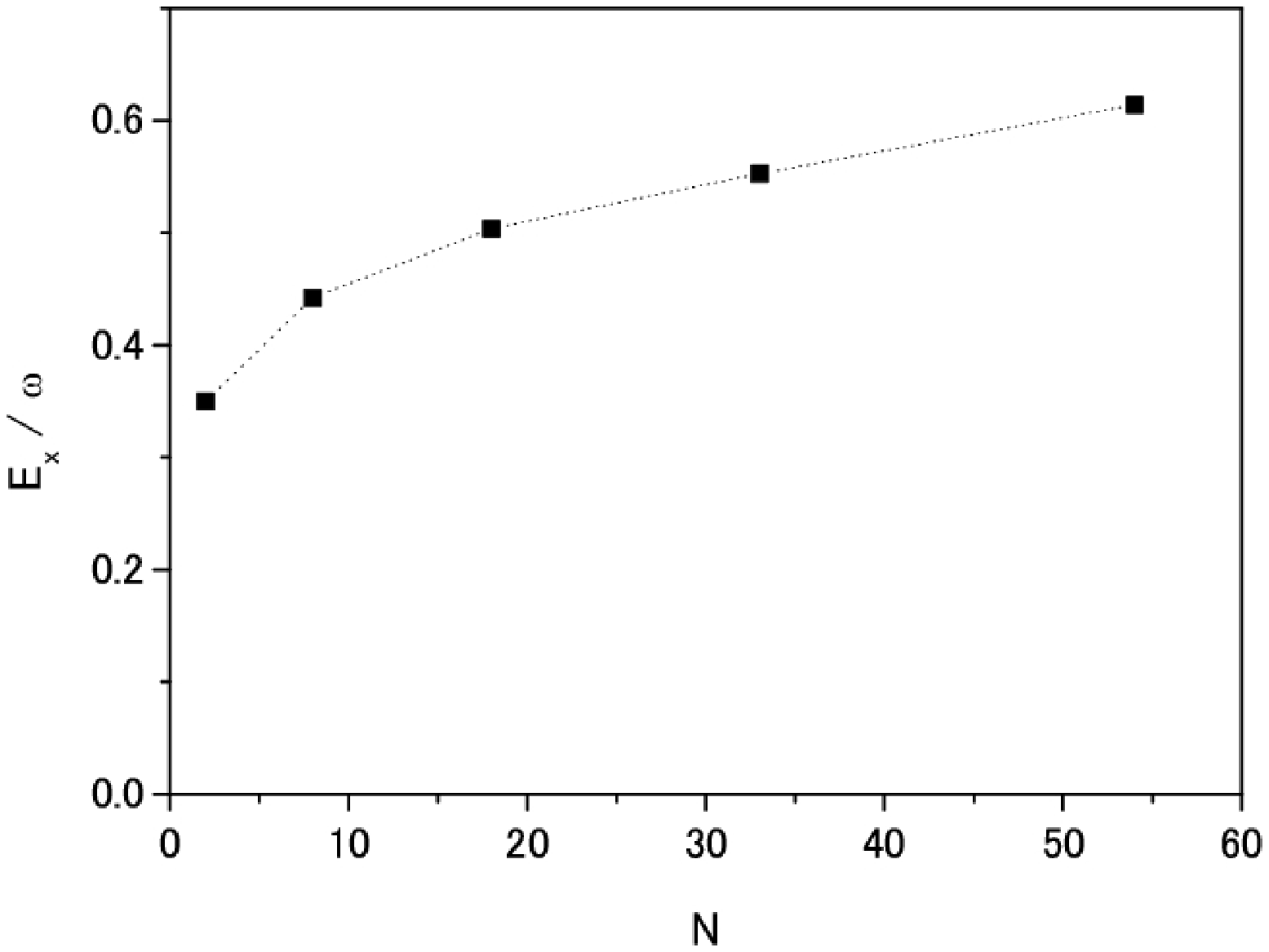}
\end{center}
\caption{Excitation energy of the scissors mode at $\chi=1$ as a function of $N$.} 
\label{fig6} 
\end{figure} 

The excitation energy is shown in Fig .\ref{fig5} as a function of $\chi$.
The fact that the excitation energy is almost
proportional to $\chi$ reflects the linear dependence of the HF energy on $\chi$ shown in Fig. \ref{fermieng}.
The excitation energy is mainly determined by the difference in the potential energy between the negative-parity states
with $|m_z|=2$ and $|m_z|=3$.
 Since the scissors mode is dominantly excited by the allowed transitions between the above-mentioned occupied 
and unoccupied single-particle states, the transition strength weakly depends on $\chi$ in contrast to the boson gas: 
The increase in the transition strength is 22\%
when $\chi$ is increased
from 0.1 to 2.
Thus the $\chi$ dependence of the energy-weighted sum is almost given by that of the excitation energy
in the case of the fermion gas considered here.

The strength function for the $yz$ quadrupole mode is shown in Fig. \ref{yz}. The small peak in the low energy region
corresponds to the scissors mode. The scissors mode of this fermion gas is energetically well separated from the $yz$ quadrupole mode,
in marked contrast to that of the boson gas.

The mass dependence of the excitation energy calculated at $\chi=1$
for the deformed configurations with $N=2$, 8, 18, 33 and 54
is shown in Fig. \ref{fig6}. For a fixed strength of the dipolar interaction the excitation energy of the 
scissors mode gradually increases with increasing $N$.

\begin{figure} 
\begin{center} 
\includegraphics[height=5cm]{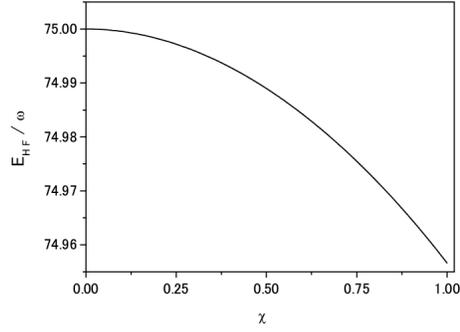}
\end{center}
\caption{Hartree-Fock energy as a function of $\chi$ for $N=20$.} 
\label{n=20} 
\end{figure} 

\begin{figure} 
\begin{center} 
\includegraphics[height=5cm]{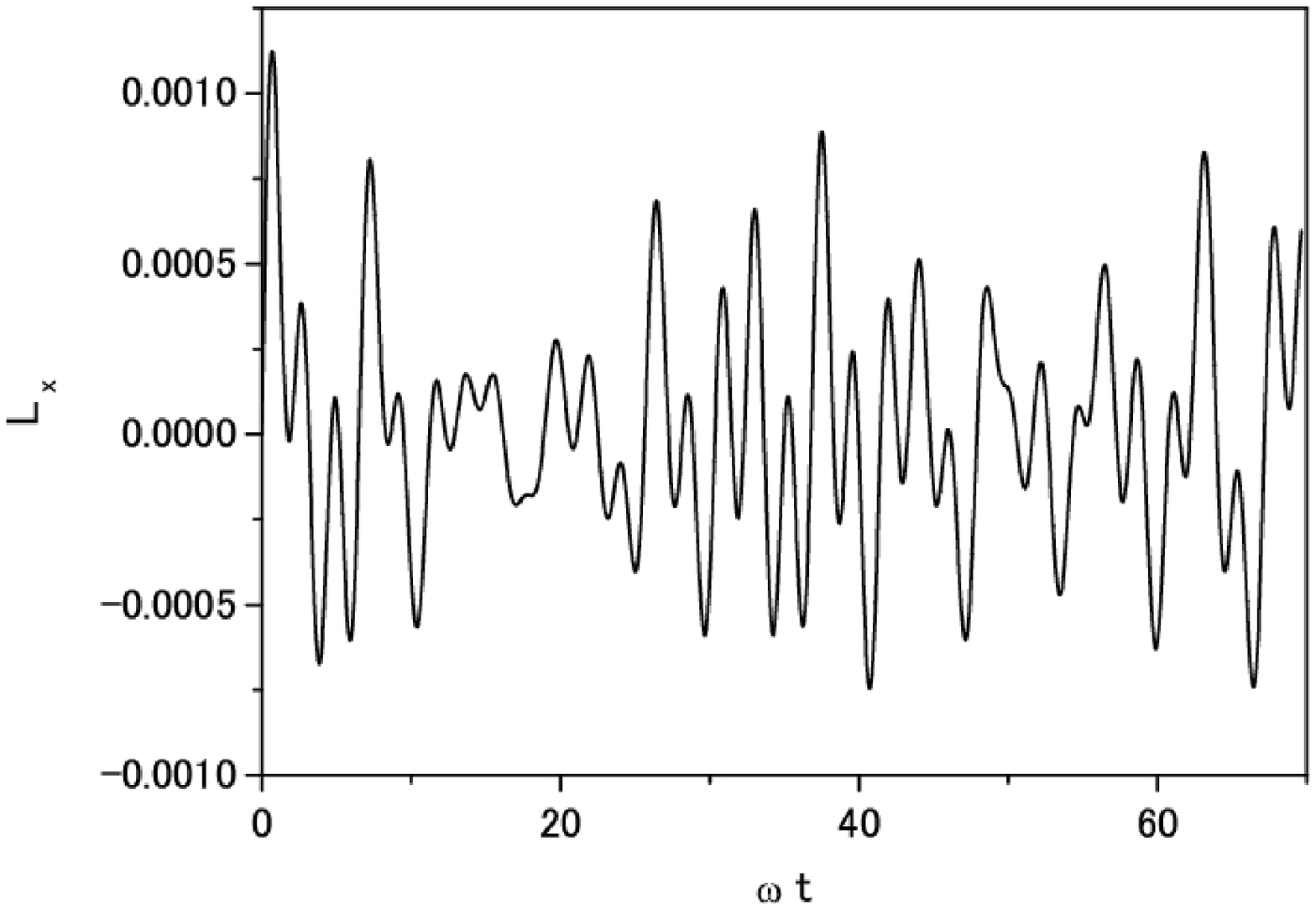}
\end{center}
\caption{Time evolution of $\langle L_x \rangle $ calculated in TDHF for $N=20$, $\chi=1$ and
$k=0.01$.} 
\label{fig4t} 
\end{figure} 

\begin{figure} 
\begin{center} 
\includegraphics[height=5cm]{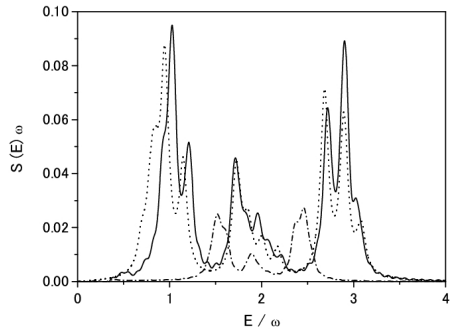}
\end{center}
\caption{Strength function calculated in TDHF for $N=20$ and $\chi=1$.
The dotted line depicts the unperturbed strength distribution. The dot-dashed line indicates the TDHF
result at $\chi=0.5$.} 
\label{fig4} 
\end{figure} 

Now we present the results for the scissors mode of a gas with a closed-shell configuration, though
it might be difficult to experimentally observe the oscillation because of a small deformation of the gas.
We consider the $N=20$ system which has the $n=3$ closed-shell configuration.
The HF energy is shown in Fig. \ref{n=20} as a function of $\chi$.
Since the attractive part and the repulsive part of the dipolar interaction are almost canceled each other,
the interaction energy is quite small. The deformation of the gas is also small \cite{Toh09}.
The HF energy decreases nearly quadratically with increasing $\chi$ 
because the energy gain is achieved through mixing of higher oscillator states as in the case of the boson gas.
The time evolution of $\langle L_x \rangle$ is shown in Fig. \ref{fig4t} for $\chi=1$. 
The strength function in TDHF (solid line) is
shown in Fig. \ref{fig4} together with the unperturbed distribution (dotted line). The samll difference between
the two calculations indicates small effects of particle - hole correlations.
The scissors mode of the close-shell gas has a broad frequency distribution centered around $2\omega$, which
reflects the degeneracy lifting within each oscillator shell due to the dipolar interaction.
The average of the excitation energies
stays around $2\omega$ independently of $\chi$, as can be seen from Fig. \ref{fig4} where the result obtained for $\chi=0.5$ is also
shown with the dot-dashed line. The fact that the average of the excitation energies is
about $2\omega$ implies that the trapping potential plays a role as a restoring force. 
The scissors mode of the closed-shell gas has small transition strengths, 
as can be understood
from the comparison of the heights of the strength functions shown in Figs. \ref{fig3} and \ref{fig4}:
The energy-weighted sum of the strength function in Fig. \ref{fig4} is only 6\% 
of that in Fig. \ref{fig3}. 
Fermion gases with the closed-shell configurations
have no allowed transitions for the scissors mode because
all the oscillators states with the same $n$ and $l$ and with different $m_z$ are almost completely occupied. 
The scissors mode can be excited
through a tiny mixing of the oscillators states caused by the dipolar interaction.
This explains the small transition strength of the scissors mode of the gas with $N=20$ . 
We also calculated 
the $yz$ quadrupole mode for $N=20$ and found that it shows an oscillation with frequency $2\omega$ as the usual
quadrupole mode excited by the operator $Y_{20}(\theta)$ \cite{Toh09}. Thus the oscillation pattern of the scissors mode
of the closed-shell gas is also distinct from
that of the $yz$ quadrupole mode.

\section{Summary}
We studied the scissors modes of dipolar boson and fermion gases trapped 
in a spherically symmetric harmonic potential. We used
the harmonic oscillator states to solve
the time-dependent Gross-Pitaevskii equation for bosons and the time-dependent Hartree-Fock equation for
fermions. It was found that the scissors mode of the dipolar boson gas shows a nearly harmonic oscillation
with quadratically increasing frequency with
increasing strength of the dipolar interaction. It was pointed out 
that both the trapping potential and the dipolar interaction act as
a restoring force of the scissors mode of the boson gas.
In the case of the dipolar fermion gas 
it was clarified that the properties of 
the scissors mode depend
on the ground-state configuration. The scissors mode of the gas with 
a deformed configuration has linearly increasing excitation energy with
increasing strength of the dipolar interaction, reflecting a pure effect of
the dipolar interaction as a restoring force. On the other hand the scissors mode of the gas
with a closed-shell configuration
shows a complicated oscillation pattern and has a broad frequency distribution centered around twice
the frequency of the trapping potential, indicating that the trapping potential also plays a role as a restoring force,
similarly to the boson gas.
It was also pointed out that the scissors mode of the dipolar fermion gas can be well distinguished from the $yz$ (or $xz$) 
quadrupole mode, in contrast to that of the boson gas which has the same excitation energy as the $yz$ (or $xz$) quadrupole mode. 
Our study was restricted 
to fermion gases with simple ground-state configurations. It is an interesting problem to
investigate the scissors mode 
of fermion gases consisting of multiple configurations, where the effects of ground-state correlations may 
have to be considered.

\end{document}